\documentclass[english,showpacs,aps,pra,reprint,amsmath,amssymb]{revtex4-1}
\usepackage[T1]{fontenc}
\usepackage[latin9]{luainputenc}
\usepackage{babel}
\usepackage{amsmath}
\usepackage{amssymb}
\usepackage{graphicx}
\usepackage[unicode=true,pdfusetitle,
 bookmarks=true,bookmarksnumbered=false,bookmarksopen=false,
 breaklinks=false,pdfborder={0 0 1},backref=false,colorlinks=false]
 {hyperref}

\makeatother

\begin{document}

\title{High-dimensional temporal mode propagation in a turbulent environment}

\author{Quanzhen Ding}

\author{Rupak Chatterjee}

\author{Yuping Huang}

\author{Ting Yu}
\email{Corresponding author: ting.yu@stevens.edu}

\affiliation{Department of Physics, and Center for Quantum Science and Engineering, 
Stevens Institute of Technology~\\
Hoboken, New Jersey 07030, USA}

\date{July 2, 2019}
\begin{abstract}
Temporal modes of photonic quantum states provide a new framework
to develop a robust free-space quantum key distribution (QKD) scheme
in a maritime environment. We show that the high-dimensional temporal
modes can be used to fulfill a persistent communication channel to
achieve high photon-efficiency even in severe weather conditions.
We identify the parameter regimes that allow for a high-fidelity quantum
information transmission. We also examine how the turbulent environment
affects fidelity and entanglement degree in various environmental
settings. 
\end{abstract}
\maketitle

\section{INTRODUCTION}

The past decades have seen a great deal of interest in photonic quantum
communication \cite{key-11-1} from several motivations including
the security of the quantum key distribution (QKD) in the presence
of environmental noise \cite{key-1,key-2,key-2-1}, the persistence
of entanglement in terms of Bell inequality violation \cite{key-3},
quantum network \cite{key-4}, and the transfer between flying qubits
(photons) and stationary qubits (atoms etc.) \cite{key-5,key-6}.
From a more fundamental physics viewpoint, the free-space quantum communication
offers a great potential of distributing information via photons promising,
in principle, unconditional security \cite{key-7,key-8,key-9,key-10}.
In practice, however, it has been shown in many realistic contexts
for deployment that the errors caused by noises or imperfections remain
the major obstacles in implementing a reliable long-distance quantum
channel when the environmental factors are not taken into account
properly~\cite{key-2,key-11,key-12}.

Several prominent quantum communication protocols have been focused
on exploring the aspects of photonic variables that allow efficient
quantum state preparations, manipulations and detections \cite{key-1,key-3,key-15,key-16,key-17,key-18,key-19-1,key-20-1,key-1322,key-35,key-36,key-36-1,key-4333,key-73-1,key-19}.
For example, the photon polarizations have been widely used in implementing
quantum key distribution (QKD) scheme in many physical settings \cite{key-1,key-3}.
However, apart from the well-known issues associated with the photon
generation, detection efficiency, and polarization imperfections, it also
becomes clear that the low-dimensional states formed by photon polarizations
may be vulnerable to the effect of environmental noises. Hence, it may
impose a limit on the capacity of the quantum communication channels.
As such, recent research has extended to the high-dimensional spaces
exploiting the advantages of multi-state systems generally called
qudits. In this context, several other types of degrees of freedom
of a single photon have been used to encode information including
orbital angular momentum (OAM) \cite{key-15,key-16,key-17,key-18,key-19-1,key-20-1},
momentum-position\cite{key-1322}, temporal modes \cite{key-35,key-36,key-36-1},
and time-energy \cite{key-4333,key-73-1,key-19}, to name a few. While
the qudit systems offer advantages in the photon information capacity
with high levels of security, in practical implementations, there
are still some shortcomings to be overcome such as the reliable OAM
state generation and the perturbation sensitivity of scattering and
absorption.

A maritime environment is notorious for causing various errors due
to scattering, absorption, and turbulence. It is imperative that the
focus of free-space quantum communication be broadened to examine
the feasibility of implementing high-dimensional photon states, which
are known to be sensitive to the atmospheric perturbation such as
turbulence. In general, the vulnerability of the photons in different
frequency domains to the turbulence destruction still remain under
investigated. Most theoretical studies engaging photon free-space
communication are usually conducted within the framework of weak scintillation,
which cannot be completely reliable if the near-sea-level implementation
of the QKD scheme is needed. Several approaches have been studied
to overcome the reduction of implementation feasibility of photon
communication channel for high-dimensional angular OAM states in a
maritime environment, a focus on a systematic investigation on the
temporal photonic mode propagation in a turbulent scenario is clearly
desirable.

In this paper, we will consider free-space temporal mode propagation
encoded by either a single photon or an entangled photon pair through
a maritime environment. Our purpose is to examine the robustness and
practicality of implementing a QKD channel in the presence of maritime
noise sources such as scattering, absorption and turbulence. In particular,
we will employ the so-called infinitesimal propagation method to simulate
the temporal mode propagation in the atmospheric setting with varied
turbulence strengths . We show by numerical simulations that the temporal
modes as the information carriers can be used to efficiently generate
quantum keys in certain frequency ranges even in the strong presence
of turbulence. We also identify some frequency ranges to be not an
ideal choice for implementing a near-sea surface quantum channel due
to the scattering and absorption processes when a near-sea surface
implementation is needed.

The structure of the paper is organized as follows. First, in Sec.~\ref{survey},
we make a brief summary of our numerical simulations on the feasibility
of the frequency ranges in the optical communication in a maritime
environment when the scattering and absorption processes are taken
into account. Such a general survey is useful in putting our investigation
into perspective, and may be used to identify the frequency ranges
that will underpin the persistent photonic communication in a maritime
environment. In Sec. III, we study the feasibility of the temporal
mode implementations under the influence of turbulence. We
conclude in Sec. IV., while some technical details are left in Appendix
A.

\section{Photonic communication in a generic maritime environment}

\label{survey}To begin, we first investigate through numerical simulations
how a maritime environment affects the photon propagation in various
parameter regimes. For a free-space optical communication system,
there are many atmospheric factors that can significantly impact photon
propagation including scattering, absorption and turbulence. The previous
work on studying the free-space optical communication channels has
shown that the environmental noises may affect the photon propagation
at every stage of the process, from the generation of photons, propagation,
to ultimately their detection by receivers. In particular, in the
propagation process, the loss of photons due to scattering or absorption
may reduce the efficacy of communication to such a level to make a
reliable communication impossible. Therefore, the successful implementation
of a theoretical concept of optical communication protocol requires
a deeper understanding of how the photon beams with different wavelengths
interacting with an atmospheric environment.

Such an investigative survey through numerical simulations of pulsed
photon beam propagation in the maritime atmosphere will provide an
useful picture about the environment-system interaction as a whole,
and will be the first step for pinpointing some optimal parameter
regimes that underpin reliable high efficiency free-space communication.
For this purpose, we have tested and conducted a Monte-Carlo method
to study the generic influence of atmospheric aerosols on the photon
communication in a maritime setting. For the processes involving molecule
scattering and absorption, we have used the 1976 U.S. Standard atmospheric
model \cite{key-22-1}. Moreover, we have employed Mie scattering
theory in our analysis \cite{key-27,key-28} and we have extensively
used the aerosol parameters from the Advanced Navy Aerosol Model (ANAM)
\cite{key-25-1,key-26}.

\begin{figure}[h]
\includegraphics[scale=0.68]{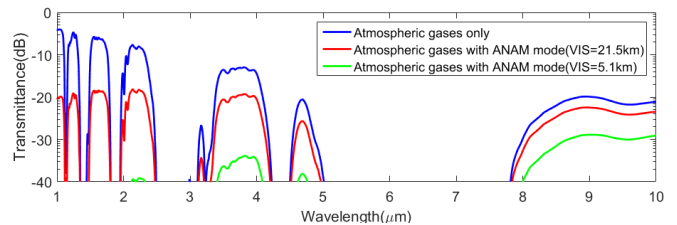}

\caption{\label{fig:epsart-1} Near center transmittance of Gaussian beams
with 10km Rayleigh length. Red: air mass parameter $p=6$, wind speed
$w=10\ \text{m/s}$, height $h=20\ \text{m}$, relative humidity $s=0.8$.
Green: air mass parameter $p=10$, wind speed $w=10\ \text{m/s}$,
height $h=19\ \text{m}$, relative humidity $s=0.95$.}

\end{figure}

Without loss of generality, we assume that the heights of the transmitter
and receiver are both at $19\ \text{m}$ for a $30\ \text{km}$ propagation
range, the photon beams emitted at this height have at least $1.36\ \text{m}$
clearance from the ocean surface, therefore, the photon beams are,
by and large, well above the ocean waves in their propagation processes.
Since the environmental impacts caused by atmospheric extinction between
the transmitter and receiver may have very large variations, it is
desirable to describe the dynamical processes beyond a naive statistical
average approach. Moreover, in order to make our numerical simulations
more dependable for the implementation of real-world quantum communication
protocols, we have considered a wide parameter range including high
relativity humidity, large air mass parameters and large wind speeds
to reflect the severe weather conditions. The major findings based
on our numerical simulations are summarized in Fig.~(\ref{fig:epsart-1}),
which exhibits a paradigmatic case involving a Gaussian beam with
$10\ \text{km}$ Rayleigh range, and the assumption that the receiver
is placed at $z_{f}=30\ \text{km}$. Among many informative results,
the simulations have indicated that the wavelength at $3.95\ \text{\ensuremath{\mu}m}$
is a desirable choice for the implementation of maritime optical communication. The numerical survey has
paved the way for further investigations of the QKD implementation
in a maritime environment. It should be emphasized, excluding some
wavelengths that are known to prone to the water absorption, that
there exist other wavelength ranges (e.g., $9-10\ \mu m$) that might
be equally feasible for the maritime QKD realization. However, the
lack of a reliable long-wave length single photon source and the detection
inefficiency are hindering their applications. Throughout our numerical
simulations, we have selected $\frac{\omega_{p}}{2}=\frac{2\pi c}{\lambda}\approx477\ \text{THz}$
as our center frequency.

\section{Photon propagation in a turbulence environment}

\subsection{Schmidt decomposition of bi-photon states}

For implementing a quantum communication protocol, let us now consider,
at the transmitter part, a two-photon state, generated by a nonlinear
crystal, represented by a Schmidt decomposition. This situation under
consideration is to assume that the down-converted beams are constrained
to be collinear with the pump beams through the use of pinholes. More
explicitly, a double-Gaussian spectral state may be written as \cite{key-19},
\begin{align}
|\Psi\rangle & =\sqrt{\frac{2}{\pi\sigma_{a}\sigma_{b}}}\int d\omega_{1}\int d\omega_{2}\exp\left[-\frac{(\omega_{1}+\omega_{2}-\omega_{P})^{2}}{2\sigma_{a}^{2}}\right]\nonumber \\
 & \quad\cdot\exp\left[-\frac{(\omega_{1}-\omega_{2})^{2}}{2\sigma_{b}^{2}}\right]|\omega_{1}\rangle|\omega_{2}\rangle
\end{align}
where $\sigma_{a}$ is determined by the coherence time of the pump
field, and $\sigma_{b}$ is determined by the phase matching bandwidth
of the spontaneous parametric down-conversion (SPDC) source \cite{key-20}.
For a typical SPDC source, the range of $\sigma_{b}$ is on the order
of hundreds of MHz to several hundreds THz. Here, for our applications,
the spatial mode of each frequency can be treated as a Laguerre-Gaussian
mode with the lowest radial and azimuthal indices (known as Gaussian
beams) \cite{key-21}.

\begin{figure}[h]
\includegraphics[scale=0.35]{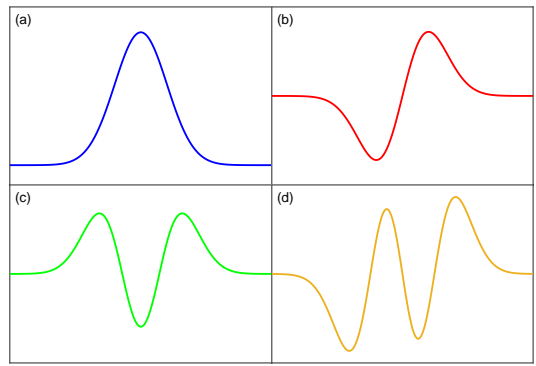}\caption{\label{fig:epsart-2}First four Schmidt mode coefficients. Blue: $f_{0}$.
Red: $f_{1}$. Green: $f_{2}$. Yellow: $f_{3}$.}
\end{figure}

For the above double-Gaussian state, an analytical expression for
the Schmidt decomposition is known as \cite{key-22,key-23}, 
\begin{align}
|\Psi\rangle & =\sum_{n=0}^{\infty}\sqrt{\lambda_{n}}|f_{n}\rangle|f_{n}\rangle\text{,}
\end{align}
with the eigenvalues $\lambda_{n}$ and the corresponding eigenstates
$|f_{n}\rangle$, 
\begin{align}
\lambda_{n} & =\frac{4\sigma_{a}\sigma_{b}(\sigma_{a}-\sigma_{b})^{2n}}{(\sigma_{a}+\sigma_{b})^{2(n+1)}},\\
|f_{n}\rangle & \equiv\int d\omega f_{n}(\omega)|\omega\rangle\\
 & =\int d\omega\left(\frac{2}{\sigma_{a}\sigma_{b}}\right)^{\frac{1}{4}}\left(2^{n}n!\sqrt{\pi}\right)^{-\frac{1}{2}}\exp\left[-\frac{(\omega-\frac{\omega_{P}}{2})^{2}}{\sigma_{a}\sigma_{b}}\right]\nonumber \\
 & \quad\cdot H_{n}\left(\sqrt{\frac{2}{\sigma_{a}\sigma_{b}}}(\omega-\frac{\omega_{P}}{2})\right)|\omega\rangle,
\end{align}
where $H_{n}(x)$ is the $n$th Hermite polynomial of $x$. Note that
$\lambda_{n}\rightarrow0$ very quickly for a large $n$. Setting
$b=\frac{2}{\sigma_{a}\sigma_{b}}$ and $\tilde{\omega}=\omega-\frac{\omega_{p}}{2}$,
as an illustration, the first four modes are given by, 
\begin{align}
|f_{0}\rangle & =\int d\omega\left(\frac{b}{\pi}\right)^{\frac{1}{4}}e^{-\frac{b}{2}\tilde{\omega}^{2}}|\omega\rangle,\\
|f_{1}\rangle & =\int d\omega\left(\frac{4b^{3}}{\pi}\right)^{\frac{1}{4}}e^{-\frac{b}{2}\tilde{\omega}^{2}}\tilde{\omega}|\omega\rangle,\\
|f_{2}\rangle & =\int d\omega\left(\frac{b}{64\pi}\right)^{\frac{1}{4}}e^{-\frac{b}{2}\tilde{\omega}^{2}}(4b\tilde{\omega}^{2}-2)|\omega\rangle,\\
|f_{3}\rangle & =\int d\omega\left(\frac{b^{3}}{768\pi}\right)^{\frac{1}{4}}e^{-\frac{b}{2}\tilde{\omega}^{2}}(8b\tilde{\omega}^{3}-12\tilde{\omega})|\omega\rangle.
\end{align}
Note that the higher orders can also be obtained in a straightforward
way. As it becomes clearer later, the higher-order modes give negligible
contributions to our analysis. Fig.~\ref{fig:epsart-2} plots the
basic behaviors of these four functions.

The Schmidt number that characterizes the degree of photonic entanglement
in the state $|\Psi\rangle$ is given by 
\begin{align}
K & =\left(\sum_{n=0}^{\infty}\lambda_{n}^{2}\right)^{-1}=\frac{\sigma_{a}^{2}+\sigma_{b}^{2}}{2\sigma_{a}\sigma_{b}}.
\end{align}

In our analysis, the condition $\sigma_{a}\ll\sigma_{b}$ is generally
valid, thus the Schmidt number can be approximated as 
\begin{align}
K & \approx\frac{\sigma_{b}}{2\sigma_{a}}.
\end{align}
Clearly, the parameter $\sigma_{a}$ of the pump laser is directly
related to the number of modes giving rise to a significant contribution
to the Schmidt number.

By using the mode-selective detection on photon 1, we can collapse
photon 2 (transmitting photon) onto one of the Schmidt modes. For
a high dimensional photonic state, the information is encoded into
the mode numbers. To be more specific, in the following discussions,
we may set $\sigma_{a}=10\ \text{THz}$ and $\sigma_{b}=80\ \text{THz}$.
For our purpose, we only consider the first four modes for encoding
the information to be transmitted through a noisy channel. It can
be easily shown that the probability $\lambda_{n}$ of finding the
state to be projected to those first four modes is given by $0.395$,
$0.239$, $0.145$, and $0.087$ respectively. As such, we see that
only $13.4\%$ of modes are discarded in state generation step. The
normalized output state from SPDC is 
\begin{align}
|\Psi\rangle & \approx(1.15)\sum_{n=0}^{3}\sqrt{\lambda_{n}}|f_{n}\rangle|f_{n}\rangle,
\end{align}
Throughout this paper, in order to avoid the degeneracy, we always
prepare the initial state in the same transverse profile (Gaussian
beam), 
\begin{align}
|f_{n}\rangle & =\int d\omega f_{n}(\omega)|0_{\omega}\rangle
\end{align}
In the next subsection, we consider how an initial photonic state
evolves in a noisy environment.

\subsection{Infinitesimal propagation equation}

An infinitesimal propagation equation (IPE) method for a single photon
with a fixed wavelength passing through a turbulent media has been
studied before in the seminal work~\cite{key-29,key-30,key-31}.
Under the monochromatic approximation, an arbitrary pure state of
a single photon can be expressed as (assuming the polarization is
uniform and may be ignored) 
\begin{align}
|\psi\rangle & =\int G(\mathbf{K},z)|\mathbf{K}\rangle\frac{d^{2}K}{4\pi^{2}}.
\end{align}
We use $z$ to denote the propagation direction where $\{|\mathbf{K}\rangle\}$
is the two-dimensional momentum basis of the transverse plane, and
$G(\mathbf{K},z)$ is the two-dimensional momentum space wave function.
The inverse Fourier transform of $G(\mathbf{K})$ gives the position
space wave function $g(x,y)$ in the transverse plane.

The equation of motion of the space wave function in the media can
be written as 
\begin{align}
\nabla^{2}E(\mathbf{x})+k^{2}n^{2}(\mathbf{x})E(\mathbf{x}) & =0,\label{eqm}
\end{align}
where $\mathbf{x}$ is the three dimensional position vector, $E(\mathbf{x})$
is the scalar part of the electric field, $k=\frac{\omega}{c}$ is
the wave number and $n(\mathbf{x})$ is the index of refraction. In
general, $n$($\mathbf{x}$) is also a function of $k$. Since the
differences of $n(\mathbf{x})$ in a narrow frequency range are negligible,
we may treat the refractive index as a frequency-independent quantity.

The refractive index $n(\mathbf{x})$ in a turbulent atmosphere may
be split into two parts, 
\begin{align}
n(\mathbf{x}) & =1+\delta n(\mathbf{x}).
\end{align}
The second term $\delta n(\mathbf{x})$ is spatially dependable, and
it is typically small compared to the first term (which is one). Thus,
an approximate Helmholtz equation may be obtained from Eq.~(\ref{eqm}),
\begin{align}
\nabla^{2}E(\mathbf{x})+k^{2}E(\mathbf{x})+2k^{2}\delta n(\mathbf{x})E(\mathbf{x}) & =0.
\end{align}
Here, we have assumed that the beam is paraxial and propagates in
the z-direction. If we decompose $E(\mathbf{x})\equiv g(\mathbf{x})e^{-ikz}$,
we can get the following paraxial wave equation for $g(\mathbf{x})$,
\begin{align}
\nabla_{T}^{2}g(\mathbf{x})-2ik\partial_{z}g(\mathbf{x})+2k^{2}\delta n(\mathbf{x})E(\mathbf{x}) & =0.
\end{align}

Using the inverse two-dimensional Fourier transform 
\begin{align}
g(\mathbf{x}) & =\int G(\mathbf{K},z)e^{-i\mathbf{K}\cdot(x,y)}\frac{d^{2}K}{4\pi^{2}},
\end{align}
we obtain 
\begin{align}
\partial_{z}G(\mathbf{K},z) & =\frac{i}{2k}|\mathbf{K}|^{2}G(\mathbf{K},z)-ikN(\mathbf{K},z)\star G(\mathbf{K},z)
\end{align}
where $N(\mathbf{K},z)$ is the inverse two-dimensional Fourier transform
of $\delta(\mathbf{x})$ and $\star$ stands for the convolution product.

The density operator for a fixed wavelength single photon state can
be expressed in the orbital angular momentum (OAM) basis, 
\begin{align}
\rho & =\sum_{m,n}\rho_{m,n}|m\rangle\langle n|.
\end{align}
where $m,n$ are collective indices for both the radial ($r$) and
orbital degrees ($l$) of the Laguerre-Gaussian (LG) modes ($m=\{r_{m},l_{m}\}$)
and $|m\rangle=\int G_{m}(\mathbf{K},z)\frac{d^{2}K}{4\pi^{2}}$.

From \cite{key-29,key-30,key-31}, we get 
\begin{align}
\partial_{z}\rho_{u,v}(z) & =S_{m,u}(z)\rho_{m,v}-S_{v,n}(z)\rho_{u,n}\nonumber \\
 & \quad+L_{m,n,u,v}(z)\rho_{m,n}-L_{T}\rho_{m,n}.\label{eq:14}
\end{align}
The first two non-dissipative terms in the above equation describe
the free-space propagation, whereas the last two dissipative terms
delineate how the turbulence causes the transition among the LG modes
through scattering processes. The operator that represents the free-space
propagation without loss is given by, 
\begin{align}
S_{p,q}(z) & =\frac{i}{2k}\int|\mathbf{K}|^{2}G_{x}(\mathbf{K},z)G_{y}^{*}(\mathbf{K},z)\frac{d^{2}K}{4\pi^{2}}.\label{eq:11}
\end{align}
More explicitly, the dissipative terms of the evolution are given
by 
\begin{align}
L_{T} & =k^{2}\int\Phi_{1}(\mathbf{K})\frac{d^{2}k}{4\pi^{2}},\label{eq:12}\\
L_{m,n,u,v}(z) & =k^{2}\int\Phi_{1}(\mathbf{K})W_{m,u}(\mathbf{K},z)W_{n,v}^{*}(\mathbf{K},z)\frac{d^{2}K}{4\pi^{2}},
\end{align}
with 
\begin{align}
W_{p,q}(\mathbf{K},z) & =\int G_{p}(\mathbf{K}_{1},z)G_{q}^{*}(\mathbf{K}_{1}-\mathbf{K},z)\frac{d^{2}K_{1}}{4\pi^{2}}.\label{eq:25}
\end{align}
It should be noted, in the above equations, that the propagation vector
$\mathbf{K}=(k_{x},k_{y})$ represents the two-dimensional projection
of the three-dimensional propagation vector $\mathbf{k}=(k_{x},k_{y},k_{z})$
and the function $G_{p}(\mathbf{K},z)$ is the two-dimensional momentum
space wave function.

The formalism presented here is valid for an arbitrary power spectral
density. In Kolmogorov turbulence theory {[}citation{]} if we ignore
the effect of the inner scales and use the von Karman power spectral
density, the refractive index power spectral density $\Phi_{1}(\mathbf{K})$
can be written as, 
\begin{align}
\Phi_{1}(\mathbf{K}) & =\frac{0.033(2\pi)^{3}C_{n}^{2}(z)}{(|\mathbf{\mathbf{K}}|^{2}+\kappa_{0}^{2})^{\frac{11}{6}}},\label{eq:15}
\end{align}
where $\kappa_{0}$ is the lager outer scale parameter and $C_{n}^{2}(z)$
is the refractive index structure constant at point $z$.

Substituting Eq.~(\ref{eq:15}) into Eq.~(\ref{eq:12}), we get
\begin{align}
L_{T} & =(30.86)C_{n}^{2}\lambda^{2}\kappa_{0}^{-\frac{5}{3}},\\
L_{m,n,u,v}(z) & =\sum_{j_{1},j_{2}=0}^{\infty}\delta_{mu}\delta_{n\nu}L_{T}+8.1\delta_{l_{m}-l_{u},l_{n}-l_{v}}l(z)\nonumber \\
 & \quad\cdot2^{j_{1}+j_{2}}\Gamma\left[\frac{j_{1}+j_{2}}{2}-\frac{5}{6}\right]c_{m,u,j_{1}}c_{m,u,j_{2}}^{*},
\end{align}
with 
\begin{align}
l(z) & =C_{n}^{2}(z)\lambda^{-2}w_{0}^{\frac{5}{3}}\left(1+(\frac{\lambda z}{\pi w_{0}^{2}})^{2}\right)^{\frac{5}{6}}\\
 & \propto\begin{cases}
w_{0}^{\frac{5}{3}}\lambda^{-2} & z\ll z_{R}=\frac{\pi w_{0}^{2}}{\lambda}\\
w_{0}^{-\frac{5}{3}}\lambda^{-\frac{1}{3}} & z\gg z_{R}=\frac{\pi w_{0}^{2}}{\lambda}
\end{cases}.
\end{align}
where the relation $\lambda=\frac{2\pi}{k}$ is used. $c_{m,u,j_{1}}$
are the coefficients (for explicit expressions, see Appendix A). It
is easy to check that $L_{T}$ diverges for large outer scales ($\kappa_{0}\rightarrow0$).
Note that $L_{m,n,u,v}(z)$ contains a counter term to cancel $L_{T}$.
Besides, the non-zero terms of $L_{m,n,u,v}(z)$ exist only when $l_{m}-l_{u}-l_{n}+l_{v}=0$.

In \cite{key-29,key-30,key-31}, Roux has shown that the integral in Eq.~(\ref{eq:11})
is non-zero only if the azimuthal indices are equal and the radial
indices differ at most by one, 
\begin{align}
S_{m,n}(z) & =\begin{cases}
\frac{i(1+|l|+2r)}{2z_{R}} & \begin{cases}
l_{m}=l_{n}=l\\
r_{m}=r_{n}=r
\end{cases}\\
\frac{i(1+|l|+r)^{\frac{1}{2}}(1+r)^{\frac{1}{2}}}{z_{R}} & \begin{cases}
l_{m}=l_{n}=l\\
|r_{m}-r_{n}|=1\\
r=min\{r_{m},r_{n}\}
\end{cases}\\
0 & {\rm otherwise},
\end{cases}\label{eq:21}
\end{align}
where $z_{R}$ is the Rayleigh range of the LG modes. $r_{m}$ and
$l_{m}$ are the radial and orbital degrees of $m$th LG mode, respectively.

From Eq.~(\ref{eq:14}) we can see that high order modes will be
involved even if the initial state only contains one lowest mode (Gaussian
beam) and $\delta n(\mathbf{\text{x}})=0$ (without scintillation).
The reason is that Roux treated $\partial_{z}\rho_{u,v}(z)=\text{Tr}[(\partial_{z}\rho(z))|u\rangle\langle v|]$
for transverse planes around $z=z_{0}$ and kept $|m\rangle=\int G_{m}(\mathbf{K},z_{0})\frac{d^{2}K}{4\pi^{2}}$
unchanged. In order to simplify the process, we treat $\partial_{z}\rho_{u,v}(z)=\partial_{z}\text{Tr}[\rho(z)|u\rangle\langle v|]$
and keep $z$-dependent basis, the first terms in Eq.~(\ref{eq:14})
vanish. New IPE can be written as 
\begin{align}
\partial_{z}\rho_{u,v}(z) & =L_{m,n,u,v}\rho_{m,n}-L_{T}\rho_{m,n}.\label{eq:28}
\end{align}

If we reorganize the density matrix $\rho$ into a column vector $\rho^{r}$
by transposing every rows and merging them one by one, we can have
\begin{align}
\partial_{z}\rho^{r} & =R\rho^{r}.
\end{align}
The elements of the new matrix $R$ is obtained by $L_{m,n,u,v}$
and $L_{T}$.

\begin{figure}
\includegraphics[scale=0.65]{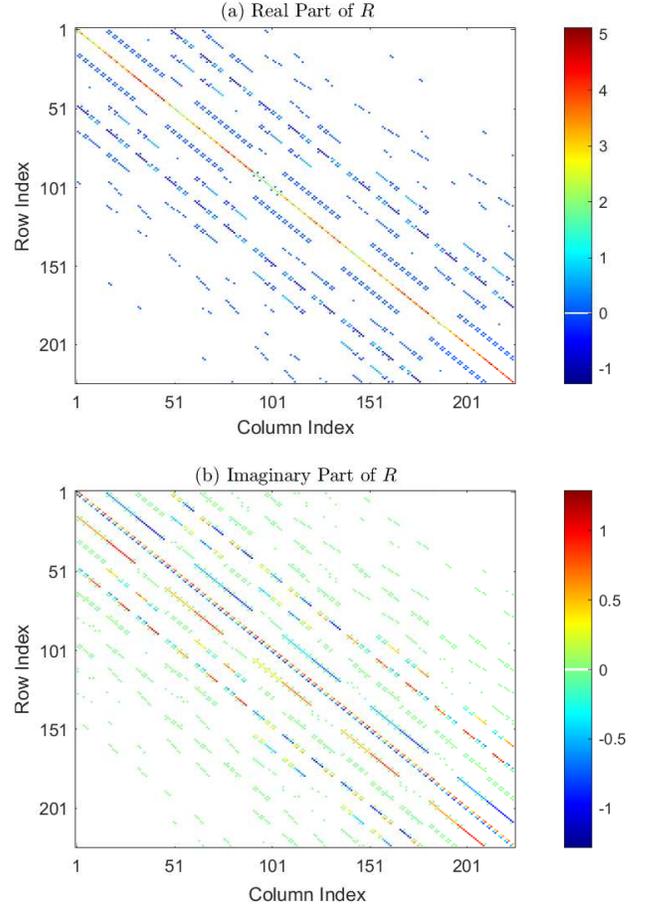}\caption{\label{fig:3}Example of the operator $R$ when $z=z_{R}$. All azimuthal
indices go from $-2$ to $2$ and all radical indices go from $0$
to $2$. Here we set $L_{0,0,0,0}-L_{T}=1$. (Jet colormap)}
\end{figure}

\begin{figure}
\includegraphics[scale=0.55]{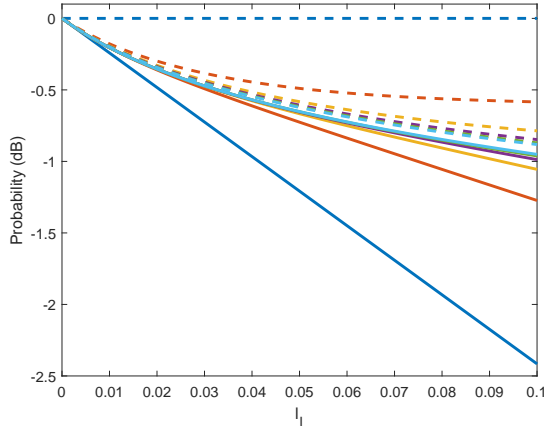}

\caption{\label{fig:4}Example of a Gaussian beam with a wavelength $\lambda=3.95\ \text{\ensuremath{\mu}m}$ and Rayleigh length $z_{R}=10\ \text{km}$
propagating in a turbulent medium. $Y$ axis means the probability
of finding the fundamental Gaussian mode while $l_{I}=\int_{0}^{z_{f}}d_{z}\ l(z)$
is a combination of multi parameters, including refractive index structure
constant, distance, wavelength (waist size can be calculated). All
azimuthal indices go from $-N$ to $N$ and all radical indices go
from $0$ to $N$. For solid lines, we adapt the cut-off of Eq. (\ref{eq:28}),
where $N$ goes from $0$ to $5$ from the bottom to the top; For
dash lines, we use the cut-off of Eq.~(\ref{eq:36}), where $N$
goes from $0$ to $5$ from the top to the bottom.}
\end{figure}

In general, Eq.~(\ref{eq:28}) represents a set of coupled first-order
differential equations. The couplings allow transitions between two
different modes. So even when the initial state contains only a few
lower order modes, the turbulence will couple those lower order modes
to all the other modes. Truncating this set of coupled equations,
one may get rid of some couplings among the participating modes. Moreover,
our numerical results have shown that the transitions become important
when the following conditions hold, 
\begin{equation}
\begin{cases}
l_{m}-l_{u}=l_{n}-l_{v}=0,\ |r_{m}-r_{u}|=|r_{n}-r_{v}|=0\\
l_{m}-l_{u}=l_{n}-l_{v}=0,\ |r_{m}-r_{u}|+|r_{n}-r_{v}|=1\\
l_{m}-l_{u}=l_{n}-l_{v}=\pm1,\ |r_{m}-r_{u}|=|r_{n}-r_{v}|=0\\
l_{m}-l_{u}=l_{n}-l_{v}=\pm1,\ |r_{m}-r_{u}|+|r_{n}-r_{v}|=1,\ 2
\end{cases}
\end{equation}
In particular, the first three cases give the most contributions.
It is interesting to know that when the azimuthal indices are unchanged,
we can barely see both $|r_{m}-r_{u}|$ and $|r_{n}-r_{v}|$ to be
equal to one. In Fig.~(\ref{fig:3}), we find that, for the OAM modes
the transitions mostly occur among their neighboring modes, and the
transition between high order and lower order modes are practically
impossible. Thus, the truncation used in IPE will contain enough useful
information about the propagation. Now we use a different Lindblad
form of IPE, written as 
\begin{align}
\partial_{z}\rho_{u,v}(z) & =L_{m,n,u,v}\rho_{m,n}-\frac{1}{2}L_{n,n,m,u}\rho_{m,v}\nonumber \\
 & \quad-\frac{1}{2}L_{m,m,v,n}\rho_{u,n}.\label{eq:36}
\end{align}
Eq.~(\ref{eq:28}) and Eq.~(\ref{eq:36}) are exactly equal since
$\sum_{n}L_{n,n,m,u}=\delta_{m,u}L_{T},\ \sum_{m}L_{m,m,v,n}=\delta_{v,n}L_{T}$.
In Fig.~(\ref{fig:4}), we plot the transition probabilities for
the two different cut-off approximations. It can be seen that the
two methods converge with increasing cut-off orders, indicating that
both methods are consistent in giving good approximate solutions.
More specifically, we set the cut off at $N=4$ for Eq.~(\ref{eq:28}).

The density operator of an general spectro-temporal single photon
in the OAM basis can be expressed as, 
\begin{align}
\rho & =\int\int d\omega_{1}d\omega_{2}\Sigma_{m_{\omega_{1}},n_{\omega_{2}}}\rho_{m_{\omega_{1}},n_{\omega_{2}}}|m_{\omega_{1}}\rangle\langle n_{\omega_{2}}|,\label{eq:8}
\end{align}
where $m_{\omega_{j}},u_{\omega_{j}}$ are collective indices for
both the radial and orbital degrees of Laguerre-Gaussian (LG) modes
with frequency $\omega_{j}$($k_{j}$). Then the equation of motion
is given by, 
\begin{align}
\partial_{z}\rho_{u_{\omega_{1}},v_{\omega_{2}}}(z) & =L_{m_{\omega_{1}},n_{\omega_{2}},u_{\omega_{1}},v_{\omega_{2}}}(z)\rho_{m_{\omega_{1}},n_{\omega_{2}}}\nonumber \\
 & \quad-(L_{T})_{\omega_{1},\omega_{2}}\rho_{u_{\omega_{1}},v_{\omega_{2}}},
\end{align}
with 
\begin{align}
(L_{T})_{\omega_{1},\omega_{2}} & =k_{1}k_{2}\int\Phi_{1}(\mathbf{K})\frac{d^{2}K}{4\pi^{2}},\\
L_{m_{\omega_{1}},n_{\omega_{2}},u_{\omega_{1}},v_{\omega_{2}}}(z) & =k_{1}k_{2}\int\Phi_{1}(\mathbf{K})W_{m_{\omega_{1}},u_{\omega_{1}}}(\mathbf{K},z)\nonumber \\
 & \quad\cdot W_{n_{\omega_{2}},v_{\omega_{2}}}^{*}(\mathbf{K},z)\frac{d^{2}K}{4\pi^{2}}.\label{eq:31}
\end{align}
We note that more general cases can be dealt with similarly the above
results after some necessary modifications.

\subsection{Gaussian beam propagation in turbulence}

We will assume that the initial state is a Gaussian beam $|0_{\omega}\rangle$
($r=0,l=0$, lowest LG mode) with a fixed wavelength $\omega$. When the high order
modes are ignored in the propagation process, the equation of motion
is given by, 
\begin{align}
\partial_{z}\rho_{u,v}(z)= & \sum_{m,n,u,v=0}^{3}L_{m,n,u,v}\rho_{m,n}-\sum_{m,n=0}^{3}L_{T}\rho_{m,n}.\label{emotion}
\end{align}

In the limit of small $\kappa_{0}$, one finds that the most essential
term is given by 
\begin{align}
L_{0,0,0,0}(z) & \approx L_{T}-(54.11)l(z)
\end{align}

We define a probability
of finding a photon in the lowest LG mode at position $z$ as
\begin{align}
P(z) &= \rho_{0,0}(z).\label{prob}
\end{align}

Having established the equation of motion (\ref{emotion}), it is
easy to show that the truncated density matrix gives rise to a pure
decay for weak turbulence or short distances, and the probability
in this regime is given by, 
\begin{align}
P(z) & =e^{-(54.11)\int_{0}^{z}l(z')dz'}
\end{align}

\begin{figure}
\includegraphics[scale=0.45]{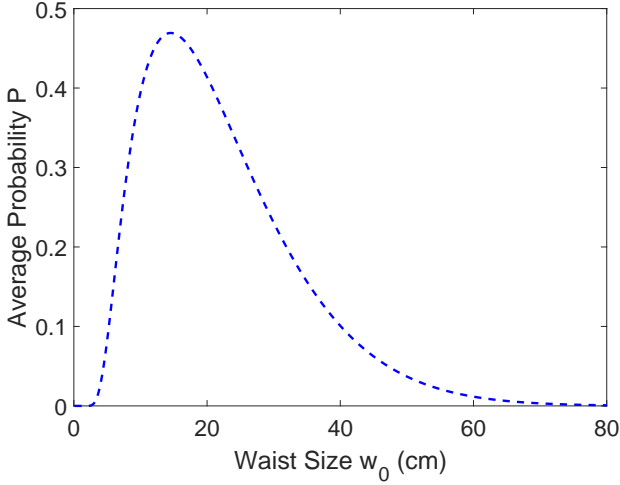}

\caption{\label{fig:epsart2}Average probability $P$ at receive plane. Distance
$\text{z=30\ km}$, wavelength $\lambda=3.95\ \text{\ensuremath{\mu}m}$,
and the refractive index structure constant $C_{n}^{2}=10^{-16}\ \text{m}^{-2/3}$.
The peak occurs at around $w_{0}=\frac{3}{4}(\frac{\lambda z}{\pi})^{1/2}\approx14.57\ \text{cm}$.}
\end{figure}

Note that the strength of the scintillation is determined by $C_{n}^{2}$
with values ranging from $10^{-13}\ \text{m}^{-2/3}$ for strong turbulence
to $10^{-17}\ \text{m}^{-2/3}$ for weak turbulence. To begin with,
we first assume that the refractive index structure constant $C_{n}^{2}$
does not vary along the propagation path $z$. Then, from
the analytical expression of $l(z)$, one can show that, for a given
waist size, the average probability monotonically increases when the
wavelength $\lambda$ increases. This observation is always valid
independent of the propagation distance. For a fixed wavelength and
propagation distance, we have shown that the minimum value of $l(z)$
occurs when $w_{0}=(\frac{\lambda z}{\pi})^{\frac{1}{2}}$ and numerical
simulations show that the maximum value of probability $P$ occurs
at around the value $w_{0}=\frac{3}{4}(\frac{\lambda z}{\pi})^{\frac{1}{2}}$
(see Fig.$\ $(\ref{fig:epsart2})). It should be noticed that it
is important to understand these dynamic behaviors when we consider
the superposition of Schmidt modes, rather than one single transition
mode.

When the propagation distance is significantly shorter than the Rayleigh
range, the probability (\ref{prob}) may be further approximated by, 
\begin{align}
P & \approx e^{-(3.25)\left(\frac{w_{0}}{r_{0}}\right)^{\frac{5}{3}}},
\end{align}
where so-called Fried parameter $r_{0}=0.185\left(\frac{\lambda^{2}}{C_{n}^{2}z}\right)^{\frac{3}{5}}$.

\begin{figure}
\includegraphics[scale=0.6]{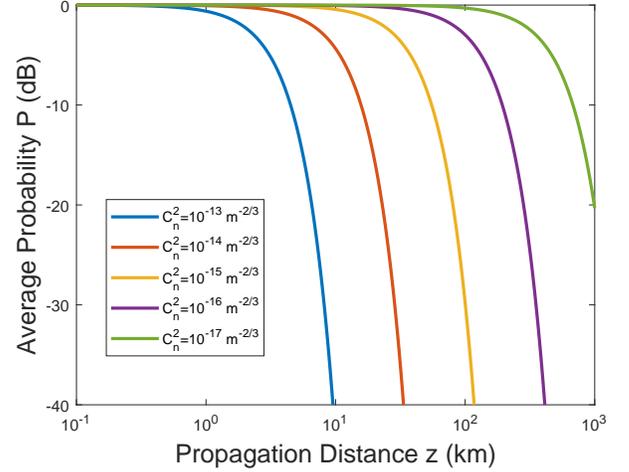}\caption{\label{fig:6-1}Average probability P of receiving the lowest LG mode versus
propagation distance. Wavelength $\lambda=3.95\ \text{\ensuremath{\mu}m}$,waist size $w_{0}=\frac{3}{4}(\frac{\lambda z}{\pi})^{1/2}$.
From left to right $C_{n}^{2}=\{10^{-13},10^{-14},10^{-15},10^{-16},10^{-17}\}\ \text{m}^{-2/3}$.}
\end{figure}

The average probability $P$ of finding photons at the position $z$
is shown to be an exponentially decaying function of the refractive
index structure constant $C_{n}^{2}$. Therefore, it is difficult
to realize the long range communication in the strong scintillation
regime (see Fig.$\ $(\ref{fig:6-1})).

\begin{figure}
\includegraphics[scale=0.45]{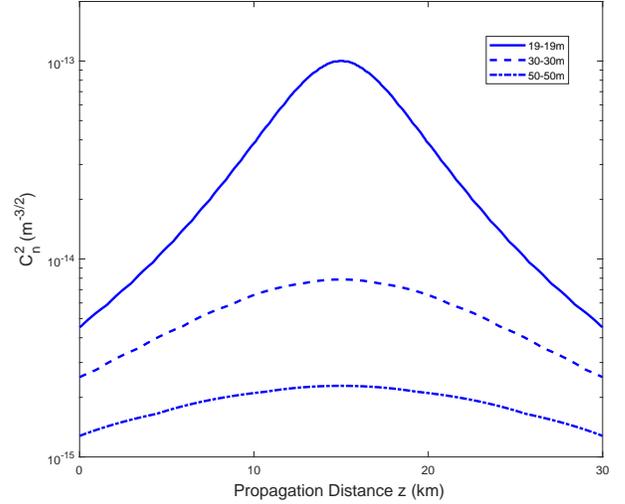}

\caption{\label{fig:epsart4}Refractive Index Structure Constants along the
propagation path in an extreme bad weather condition. Sea level temperature
$T_{sea}=15\protect\textdegree C$, air temperature $T_{air}=20\protect\textdegree C$,
wind speed $w=5m/s$, and relative humidity $RH=98\%$. Solid line:
$19-19\ \text{m}$; dash line: $30-30\ \text{m}$; dash-dot line:
$50-50\ \text{m}$. Earth curvature is considered.}
\end{figure}

In general, $C_{n}^{2}(z)$ is highly sensitive to the propagation
path height, but is relatively insensitive to the wavelength. Our
simulations for the turbulence are based on varieties of parameter
ranges to better reflect a real maritime environment.

It can be shown that when the wavelength is greater than $2\ \text{\ensuremath{\mu}m}$,
the refractive index structure constant is no longer dependent on
the wavelength \cite{key-32,key-33}. Our simulations investigate
the bad weather condition represented by high relative humidity, strong
wind speed and large air-sea surface temperature difference.
Turbulence typically falls off sharply in the first few meters above
the sea surface, then it varies slowly with the increasing height.
To have a more accurate description of the long distance maritime
communication, the Earth curvature should be taken into account. In
Fig.~(\ref{fig:epsart4}), we have plotted three cases about the
propagation path: the heights of the emitter and receiver are $19-19\ \text{m}$,
$30-30\ \text{m}$, and $50-50\ \text{m}$, respectively. The results
show that the strong turbulence may be avoided when the emitter and
receiver are well above the sea surface. For example, in the $19-19\ \text{m}$
case, we see that the maximum refractive index structure constant
along the propagation path $C_{n}^{2}$ is about $2\times10^{-13}\ \text{m}^{-2/3}$,
which makes practical communication impossible. Therefore, one needs
to use other paths.

\subsection{Temporal mode propagation}

For an initial state prepared in the $n$th temporal mode, the reduced
density operator can be written as 
\begin{align}
\rho_{n}^{i} & =\int\int d\omega_{1}d\omega_{2}f_{n}(\omega_{1})f_{n}(\omega_{2})|0_{\omega_{1}}\rangle\langle0_{\omega_{2}}|,\label{eq:8-1}
\end{align}

Since the infinitesimal propagation method also works for the cross
terms $|0_{\omega_{1}}\rangle\langle0_{\omega_{2}}|$($\lambda_{j}=\frac{2\pi c}{\omega_{j}}$),
we find that the modified $l(\omega_{1},\omega_{2},z)$ takes the
following form 
\begin{align}
l(\omega_{1},\omega_{2},z) & \approx C_{n}^{2}(z)(\lambda_{1}\lambda_{2}){}^{-1}w_{0}^{\frac{5}{3}}\nonumber \\
 & \quad\cdot\left[1+\frac{1}{2}(\frac{\lambda_{1}z}{\pi w_{0}^{2}})^{2}+\frac{1}{2}(\frac{\lambda_{2}z}{\pi w_{0}^{2}})^{2}\right]^{\frac{5}{6}}.
\end{align}

The influence of scintillation will typically evolve the initial pure
state into a mixed state. For our purpose here, we only need to consider
the lowest LG mode at the receive plane. Therefore, the truncated
final density matrix at the propagation distance $z$ may be written
as, 
\begin{align}
\rho_{n}^{f} & =\int\int P(\omega_{1},\omega_{2})f_{n}(\omega_{1})f_{n}(\omega_{2})|0_{\omega_{1}}\rangle\langle0_{\omega_{2}}|d\omega_{1}d\omega_{2}.\label{temporal}
\end{align}

\begin{figure}
\includegraphics[scale=0.44]{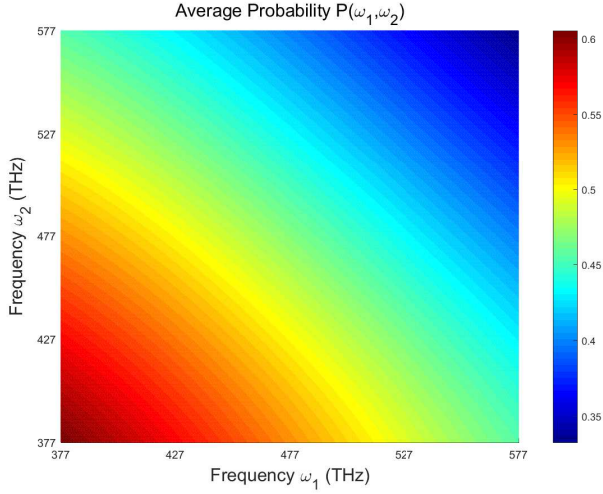} \caption{\label{fig:epsart5}Average probability $P$ for both diagonal and
cross terms. Propagation distance $z=30\ \text{km}$, waist size $w_{0}=14.57\ \text{cm}$
and the refractive index structure constant $C_{n}^{2}=10^{-16}\ \text{m}^{-2\text{/3}}$. }
\end{figure}

Note the trace of the density matrix (\ref{temporal}) denoted by
$T_{n}=\int P(\omega,\omega)|f_{n}(\omega)|^{2}d\omega$ is not 1.
The decaying function $T_{n}$ represents the information of the transition
from the $n$th mode to the other modes. The normalized density matrix
$\rho_{n}^{N}$ is obtained by dividing $T_{n}$, 
\begin{align}
\rho_{n}^{N} & =\frac{1}{T_{n}}\int\int P(\omega_{1},\omega_{2})f_{n}(\omega_{1})f_{n}(\omega_{2})|\omega_{1}\rangle\langle\omega_{2}|d\omega_{1}d\omega_{2}.
\end{align}

One can show that the probability of finding the photon at the receiver
position in the $m$th mode is 
\begin{align}
S_{n,m} & =\frac{1}{T_{n}}\int\int P(\omega_{1},\omega_{2})f_{m}(\omega_{1})f_{m}(\omega_{2})\nonumber \\
 & \quad\cdot f_{n}(\omega_{1})f_{n}(\omega_{2})|\omega_{1}\rangle\langle\omega_{2}|d\omega_{1}d\omega_{2}.
\end{align}

\begin{figure}
\includegraphics[scale=0.45]{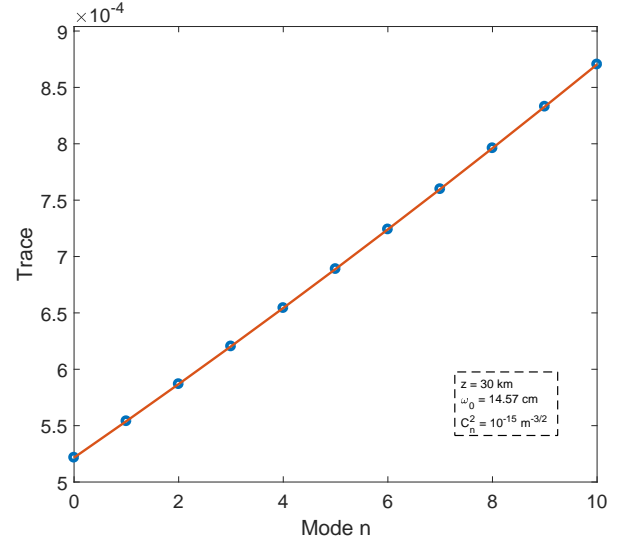}\caption{\label{fig:6}The total traces (probabilities) of the first 11 modes.}
\end{figure}

We have set the distance $\text{z=30\ km}$, wavelength $\lambda=3.95\ \text{\ensuremath{\mu}m}$,
and the refractive index structure constant $C_{n}^{2}=10^{-15}\ \text{m}^{-2/3}$
to test our approach. The total traces (probabilities of finding this
mode) of the first four modes are shown in Fig.$\ $(\ref{fig:6}).
We have shown that for this turbulence condition the transmittance
of our time modes are still in an acceptable range (< 40 dB).

More explicitly, the transition probabilities of the first four modes
are given by the following transmission matrix,
\begin{align}
S & =\begin{bmatrix}0.9838 & 0.0161 & 0.0000 & 0.0000\\
0.0152 & 0.9538 & 0.0307 & 0.0003\\
0.0001 & 0.0289 & 0.9266 & 0.0438\\
0.0000 & 0.0003 & 0.0414 & 0.9018
\end{bmatrix}.
\end{align}
Apparently, high-dimensional temporal modes can be sustained but their
stabilities become low, i.e., the probabilities of transferring to
other modes increases.

\subsection{Entangled photon propagation}

In this subsection, we will discuss the entangled photon pair propagation
in a turbulence environment. To begin, we consider two entangled photons
with the following initial state, 
\begin{align}
|\psi\rangle & =\frac{1}{\sqrt{2}}(|f_{m}\rangle|f_{m}\rangle+|f_{n}\rangle|f_{n}\rangle),
\end{align}
Then, the reduced density operator is given by, 
\begin{align}
\rho^{i} & =\frac{1}{2}\int\int\int\int[f_{m}(\omega_{1})f_{m}(\omega_{1}')+f_{n}(\omega_{1})f_{n}(\omega_{1}')]\nonumber \\
 & \quad\cdot[f_{m}(\omega_{1})f_{m}(\omega_{1}')+f_{n}(\omega_{2})f_{n}(\omega_{2}')]\nonumber \\
 & \quad\cdot|0_{\omega_{1}},0_{\omega'_{1}}\rangle\langle0_{\omega_{2}},0_{\omega'_{2}}|d\omega_{1}d\omega_{1}'d\omega_{2}d\omega_{2}'.
\end{align}

We must adjust $l(z)$ to account for the cross terms $|0_{\omega_{1}},0_{\omega'_{1}}\rangle\langle0_{\omega_{2}},0_{\omega'_{2}}|$$\ \left(\lambda_{i}=\frac{2\pi c}{\omega_{i}}\right)$
\begin{align}
l(\omega_{1},\omega_{2},\omega_{1}',\omega_{2}',z) & \approx\pi C_{n}^{2}(z)(\lambda_{1}\lambda_{2}){}^{-1}w_{0}^{-\frac{5}{3}}\nonumber \\
 & \quad\cdot\bigg\{\left[1+\frac{1}{2}(\frac{\lambda_{1}z}{\pi w_{0}^{2}})^{2}+\frac{1}{2}(\frac{\lambda_{2}z}{\pi w_{0}^{2}})^{2}\right]^{\frac{5}{6}}\nonumber \\
 & \quad+\left[1+\frac{1}{2}(\frac{\lambda'_{1}z}{\pi w_{0}^{2}})^{2}+\frac{1}{2}(\frac{\lambda'_{2}z}{\pi w_{0}^{2}})^{2}\right]^{\frac{5}{6}}\bigg\}.
\end{align}

\begin{figure}[h]
\includegraphics[scale=0.58]{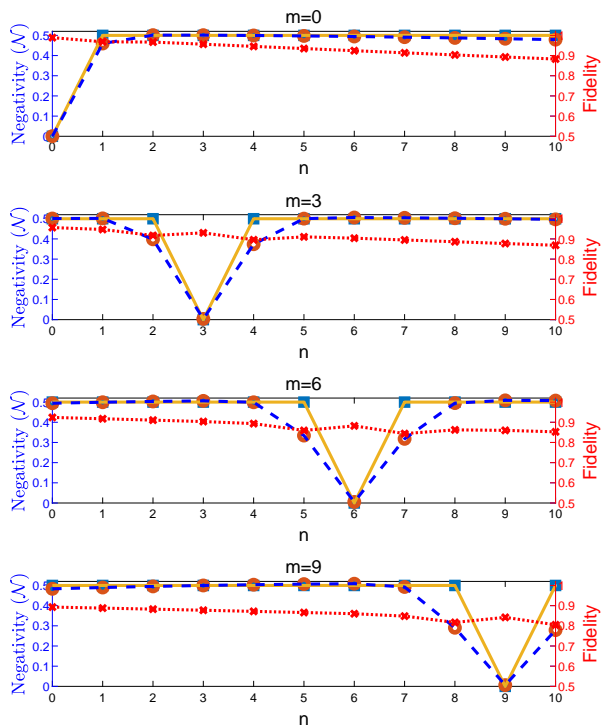}\caption{\label{entfig}Negativity and fidelity with different modes .}
\end{figure}

In order to investigate the entanglement evolution of high dimensional
entangled photon pairs in a random media, we choose the log negativity
as an entanglement measure. Fig.~(\ref{entfig}) is plotted with
the parameters $w_{0}=14.57\ \text{cm}$ and $z=30\ \text{km}$. We
consider the strong turbulence case, which is characterized by $C_{n}^{2}=10^{-16}\ \text{m}^{-2/3}$
along the propagating path. We are interested in the robustness of
entangled photon pairs under the influence of turbulence. To be more
specific, we fix the mode number $m$ for one photon and vary the
second photon's mode number $n$ from 0 to 10. The blue squares are
the negativities of the initial states and the red circles are the
log negativities of the final states at the receiving aperture. Solid
lines connect all the squares in every sub-figures where dash lines
connect the circles. The red dot lines connect all the diagonal crosses
are the fidelities between the initial states and the output states.
We can see that, when the entangled photons propagate in a strong
turbulence regime, the entanglement measured by the negativity defined
as $E_{N}(\rho)=\log_{2}[2\mathcal{N}+1]$ doesn't change too much.
It's easy to find that, when $m$ and $n$ are not close ($|m-n|>1$),
the entanglement will remain unaffected even in a strong turbulence
regime. When $|m-n|=1$, the two modes are close, interference occurs
and heavily influences the log negativity. Thus, we show that the
qudit formed by consecutive even number modes (e.g., modes 0, 2, 4)
or odd number modes (e.g, modes 1, 3, 5) will be more robust against
the random perturbations.

\section{CONCLUSION}

To summarize, based on the Schmidt decomposition representation, high
dimensional temporal mode propagation is systematically studied by
using the well-known infinitesimal propagation method. We have shown
that, in a highly dynamic maritime environment, there exist frequency
ranges that will allow the reliable implementations of photon communication
in the framework of temporal modes. In particular, we have examined
the feature of Schmidt eigenmodes and identified robust and fragile
parameter domains against log-distance extinction (scattering and
absorption). In addition, we have analysed the dynamical behavior
of entangled photon pairs under the influence of strong turbulence.
Since the colored noises and correlated noises are of importance for
free-space communication, a further study on the photonic noisy propagation would be useful  \cite{key-100,key-101}.

\section*{acknowledgements}

This research was supported in part by the Office of Naval Research
(Award No. N00014-15-1-2393). We would like to thank Drs. Yusui Chen,
Wufu Shi and Y. M. Sau for useful discussions.

\appendix

\section{Solutions of Integrals in IPE}

The purpose of this Appendix is to analyze the influence of turbulence
via infinitesimal propagation equation method. The generating function
for the LG modes of a fixed wavelength can be written as 
\begin{align}
G & =\sum_{n,m=0}^{\infty}\frac{1}{m!}L_{n}^{m}\left(\frac{2(u^{2}+v^{2})}{1+t^{2}}\right)\left[\frac{d(1+1t)}{1-it}\right]^{n}\nonumber \\
 & \quad\times\frac{[(u+iv)p+(u-iv)q]^{m}}{(1-it)^{1+m}}\nonumber \\
 & =\frac{1}{\Omega(t,d)}\exp\bigg[\frac{(u+iv)p}{\Omega(t,d)}+\frac{(u-iv)q}{\Omega(t,d)}\nonumber \\
 & \quad-\frac{(1+d)(u^{2}+v^{2})}{\Omega(t,d)}\bigg],
\end{align}
where $\Omega(t,d)=1-d-it-itd.$ The normalized coordinates are given
by $u=x/w_{0}$, $v=y/w_{0}$ and $t=z/z_{R}=z\lambda/\pi w_{0}^{2}$
in terms of the waist size $w_{0}$ at the initial location and the
Rayleigh range $z_{R}$. The parameters $p$, $q$ and $w$ are used
to generate particular LG modes in the following way,

\begin{align}
M_{r,l}^{LG}(u,v,t) & =\begin{cases}
\mathcal{N}[\frac{1}{r!}\partial_{d}^{r}\partial_{p}^{|l|}G]_{d,p,q=0} & l>0\\
\mathcal{N}[\frac{1}{r!}\partial_{d}^{r}G]_{d,p,q=0} & l=0\\
\mathcal{N}[\frac{1}{r!}\partial_{d}^{r}\partial_{q}^{|l|}G]_{d,p,q=0} & l<0
\end{cases}
\end{align}
with normalization constant 
\begin{align}
\mathcal{N} & =\left[\frac{r!2^{|l|+1}}{\pi(r+|l|)}\right]^{\frac{1}{2}},
\end{align}
where $r$ is the radial index (a non-negative integer) and $l$ is
the azimuthal index (an integer). 

In order to compute the integrals in the IPE, we must get the the
Fourier transform of the generating function, 
\begin{align}
\mathcal{F}[G] & =\frac{\pi}{1+d}\exp[\frac{i\pi(\tilde{k}_{x}+i\tilde{k}_{y})p}{1+d}\nonumber \\
 & \quad+\frac{i\pi(\tilde{k}_{x}-i\tilde{k}_{y})q}{1+d}-\frac{\pi(\tilde{k}_{x}^{2}+\tilde{k}_{y}^{2})\Omega(t,d)}{1+d}].
\end{align}

Here $\tilde{k}_{x}$ and $\tilde{k}_{y}$ are normalized spatial
frequency components that $\tilde{k}_{x}=\frac{w_{0}}{2\pi}k_{x}$
and $\tilde{k}_{y}=\frac{w_{0}}{2\pi}k_{y}$.

After evaluating the Eq.~(\ref{eq:11}) and removing the superfluous
mixed terms containing a $p$ times a $q$, one obtains a generation
function 
\begin{align}
S_{G}(z) & =\frac{i\lambda(1+d_{m})(1+d_{n})}{8(1-d_{m}d_{n})}\exp\left[\frac{p_{m}p_{n}+q_{m}q_{n}}{2(1-d_{m}d_{n})}\right]\nonumber \\
 & \quad\times[2(1-d_{m}d_{n})+p_{m}p_{n}+q_{m}q_{n}],
\end{align}
where $p_{m}$, $q_{m}$ and $d_{m}$ are generating function parameters
associated with the $m$ index, while $p_{n}$, $q_{n}$ and $d_{n}$
are generating function parameters associated with the $n$ index.
From this function, we find non-zero values occur when the azimuthal
indices involved be equal, and that the radial indices differ at most
by one, i.e., Eq.~(\ref{eq:21}).

For the term $W_{m,n}(\mathbf{K},z)$ , we must to get a generating
function for the radial indices of the modal correlation functions
\begin{align}
W_{rG}(K,\phi,z) & =\frac{\exp(-X)\exp[i(l_{m}-l_{n})\phi]E_{m}^{|l_{m}|}\bar{E}_{n}^{|l_{n}|}}{(1-d_{m}d_{n})}\nonumber \\
 & \quad\times\left[\frac{r_{m}!}{(r_{m}+|l_{m}|)!}\right]^{\frac{1}{2}}\left[\frac{r_{n}!}{(r_{n}+|l_{n}|)!}\right]^{\frac{1}{2}}\nonumber \\
 & \quad\times\sum_{s=0}^{M(l_{m},l_{n})}\frac{|l_{m}|!|l_{n}|!(-X)^{s}}{(|l_{m}|-s)!(|l_{n}|-s)!s!}
\end{align}
with 
\begin{align}
M(l_{m},l_{n}) & =\frac{1}{2}(|l_{m}|+|l_{n}|-|l_{m}-l_{n}|),\\
X & =\frac{K^{2}\zeta_{m}\zeta_{n}^{*}\eta^{2}}{8\pi^{2}(1-d_{m}d_{n})},\\
E_{m} & =\frac{iK\zeta_{m}\eta}{2\sqrt{2}\pi(1-d_{m}d_{n})},\\
\bar{E}_{n} & =\frac{iK\zeta_{n}^{*}\eta}{2\sqrt{2}\pi(1-d_{m}d_{n})}.
\end{align}

Here, we set $\zeta_{x}=z_{R}-iz-d_{x}(z_{R}+iz)$, $\eta=\frac{\lambda}{w_{0}}$,
and use we are using polar momentum space coordinates $k_{x}+ik_{y}=Ke^{i\phi}$
. Setting $a=(1+t^{2})w_{0}{}^{2},b=\frac{1+it}{1-it}$,we can get $X=\frac{K^{2}(1-d_{m}b)(1-d_{n}b^{-1})a}{8(1-d_{m}d_{n})}$, and the explicit
form of the term $W_{m,n}(K,\phi,z)$ can be written as 
\begin{widetext}
\begin{align}
W_{m,n}(K,\phi,z) & =\sum_{s=0}^{M(l_{m},l_{n})}e^{i(l_{m}-l_{n})\phi}(i)^{|lm|+|l_{n}|}(-1)^{-s}\left(\frac{K^{2}a}{8}\right)^{\frac{|l_{m}|+|l_{n}|}{2}-s}\left[\frac{1}{(r_{m}+|l_{m}|)!}\right]^{\frac{1}{2}}\left[\frac{1}{(r_{n}+|l_{n}|)!}\right]^{\frac{1}{2}}\nonumber \\
 & \quad\times\frac{|l_{m}|!|l_{n}|!}{(|l_{m}|-s)!(|l_{n}|-s)!s!}b^{-\frac{|l_{m}|-|l_{n}|}{2}}\partial_{d_{m}}^{r_{m}}\partial_{d_{n}}^{r_{n}}\left(e^{-X}\frac{(1-d_{m}b)^{|l_{m}|-s}(1-d_{n}b^{-1})^{|l_{n}|-s}}{(1-d_{m}d_{n})^{|l_{m}|+|l_{n}|-s+1}}\right)\Bigg|_{d_{m},d_{m}=0},
\end{align} 
\end{widetext}
which can also be sorted as
\begin{align}
W_{m,n}(K,\phi,z) & =\sum_{j=0}^{\infty}c_{m,n,j}\left(\frac{K^{2}a}{8}\right)^{\frac{j}{2}}e^{-\frac{K^{2}a}{8}}e^{i(l_{m}-l_{n})\phi},
\end{align}
where $c_{m,n,j}$ is a coefficient independent of $K$ and $\phi$.

When we substitute the von Karman spectrum, the two-dimensional integration
in Eq.~(\ref{eq:31}) can be split into a radial and angular integral.
Since the von Karman density $\Phi_{1}(K)$ only depends on the radial
coordinate, the integral over $\phi$ only involves the $\phi$ component
of $W_{m,n}$. Non-zero values only occur when $l_{m}-l_{u}-l_{n}+l_{v}=0$.
Setting $\kappa_{0}\rightarrow0$, we can get
\begin{widetext}
\begin{align}
L_{m,n,u,v}(z) & =k^{2}\int\Phi_{1}(\mathbf{K})W_{m,u}(\mathbf{K},z)W_{n,v}^{*}(\mathbf{K},z)\frac{d^{2}K}{4\pi^{2}}\nonumber \\
 & \approx\delta_{mu}\delta_{n\nu}L_{T}+(8.1)\delta_{l_{m}-l_{u},l_{n}-l_{v}}C_{n}^{2}\lambda^{-2}w_{0}^{\frac{5}{3}}(1+t^{2})^{\frac{5}{6}}\left\{ \sum_{j_{1},j_{2}=0}^{\infty}2^{-\frac{j_{1}+j_{2}}{2}}\Gamma\left[\frac{j_{1}+j_{2}}{2}-\frac{5}{6}\right]c_{m,u,j_{1}}c_{n,v,j_{2}}^{*}\right\} .
\end{align}
\end{widetext}

\end{document}